\documentstyle[preprint,eqsecnum,aps,floats,epsfig]{revtex}
\begin{document}
\topmargin -0.5in
\makeatletter
\def\maketitle{\par
\begingroup
\let\cite\@bylinecite
\def\thefootnote{\fnsymbol{footnote}}%
\if@twocolumn
\twocolumn[\@maketitle\vskip2pc]%
\else
\newpage
\epsfysize3cm
\epsfbox{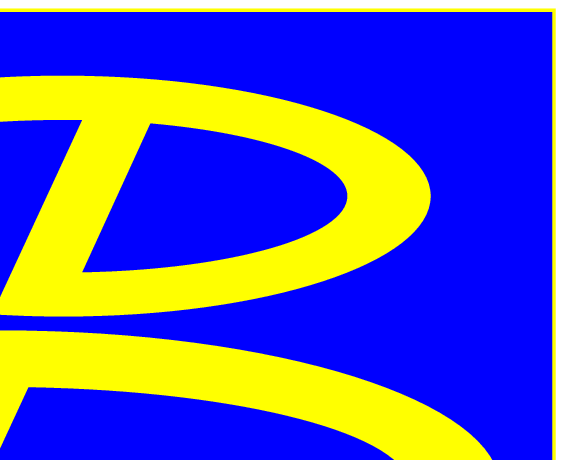}    
\global\@topnum\z@ %
\@maketitle
\fi
\thispagestyle{plain}\@thanks
\endgroup
\def\thefootnote{\arabic{footnote}}%
\setcounter{footnote}{0}%
\let\maketitle\relax \let\@maketitle\relax
\let\@thanks\relax \let\@authoraddress\relax \let\@title\relax
\let\@date\relax \let\thanks\relax
}
\makeatother
\draft
\preprint{\tighten\vbox{\hbox{\hfil KEK preprint  2000-142}
    \hbox{\hfil Belle preprint 2001-1}
    }}

\title{Measurement of the {\boldmath $CP$} Violation Parameter {\boldmath $\sin{}2\phi_1$} in {\boldmath $B^0_d$} Meson Decays}
\author{
A.~Abashian$^{44}$,
K.~Abe$^{8}$, 
K.~Abe$^{36}$, 
I.~Adachi$^{8}$, 
Byoung~Sup~Ahn$^{14}$, 
H.~Aihara$^{37}$, 
M.~Akatsu$^{19}$, 
G.~Alimonti$^{7}$, 
K.~Aoki$^{8}$,
K.~Asai$^{20}$
% ...
}

\address{
% ...
$^{7}${University of Hawaii, Honolulu HI}\\
$^{8}${High Energy Accelerator Research Organization (KEK), Tsukuba}\\
% ...
$^{14}${Korea University, Seoul}\\
% ...
$^{19}${Nagoya University, Nagoya}\\
$^{20}${Nara Women's University, Nara}\\
% ...
$^{36}${Tohoku University, Sendai}\\
$^{37}${University of Tokyo, Tokyo}\\
% ...
$^{44}${Virginia Polytechnic Institute and State University, Blacksburg VA}\\
% ...
}
\date{\today}
\maketitle
\tighten
\begin{abstract}
We present a measurement of the Standard Model $CP$ violation parameter 
$\sin 2\phi_1$ (also known as $\sin 2\beta$)  based on 
a $10.5~{\rm fb}^{-1}$ data sample collected at the $\Upsilon(4S)$ resonance
with the Belle detector at the KEKB asymmetric $e^+e^-$ collider.
One neutral $B$ meson
is reconstructed in the
$J/\psi K_S$, $\psi(2S) K_S$, $\chi_{c1} K_S$, $\eta_c K_S$, $J/\psi K_L$ 
or $J/\psi \pi^0$
$CP$-eigenstate decay channel and
the flavor of the accompanying $B$ meson is identified 
from its charged particle decay products.
From the asymmetry in the 
distribution of the time interval between the two $B$-meson decay points,
we determine 
$\sin 2\phi_1=0.58^{+0.32}_{-0.34}({\rm stat})^{+0.09}_{-0.10}({\rm syst}).$
\vskip 4cm
\center{ (submitted to Phys. Rev. Lett.)}
\vskip -4.3cm
\end{abstract}
\pacs{PACS numbers:11.30.Er,12.15.Hh,13.25.Hw}


\narrowtext
In the Standard Model (SM), $CP$ violation arises from 
a complex phase
in the Cabibbo-Kobayashi-Maskawa (CKM) quark 
mixing matrix~\cite{KM}.
In particular, the SM predicts 
a $CP$ violating asymmetry in the time-dependent 
rates for $B^0_d$ and $\overline{B_d}^0$
decays to a common $CP$ eigenstate, $f_{CP}$,
without theoretical ambiguity due to
strong interactions~\cite{carter}: 
\setcounter{section}{1}
$$\begin{array}{l}
A(t)\equiv 
\frac{\Gamma(\overline{B_d}^0\to f_{CP})-\Gamma({B^0_d}\to f_{CP})}
{\Gamma(\overline{B_d}^0\to f_{CP})+\Gamma({B^0_d}\to f_{CP})}
=-\xi_f\sin 2\phi_1 \sin\Delta m_d t,
\end{array}$$
where $\Gamma(\overline{B_d}^0$ $(B_d^0)$ 
$\to f_{CP})$ is the decay rate
for a $\overline{B_d}^0(B_d^0)$ to $f_{CP}$ at a proper time $t$ after production,
$\xi_f$ is the $CP$-eigenvalue of $f_{CP}$, 
$\Delta m_d$ is the mass difference between the two $B^0_d$ mass eigenstates, and
$\phi_1$  is one of the three internal 
angles of the CKM Unitarity Triangle, defined as
$\phi_1\equiv \pi-\arg\left(\frac{-V^*_{tb}V_{td}}{-V^*_{cb}V_{cd}}\right)$~\cite{Sanda}.

In this Letter, we report a measurement of $\sin 2\phi_1$ 
using $B^0_d\overline{B_d}^0$ meson pairs 
produced at the $\Upsilon(4S)$ resonance, where
the two mesons remain in a coherent
$p$-wave state until one of them decays.
The decay of  one of the $B$ mesons to a self-tagging state, $f_{tag}$, 
i.e.~a final state that distinguishes between $B^0_d$ and 
$\overline{B_d}^0$, at time $t_{tag}$
projects the accompanying meson onto the opposite $b$-flavor at that time;
this meson decays to $f_{CP}$ at time $t_{CP}$.
The $CP$ violation manifests itself as an asymmetry
$A(\Delta t)$,
where $\Delta t$ is the proper time interval $\Delta t\equiv t_{CP}-t_{tag}$.

The data sample corresponds to an 
integrated luminosity of $ 10.5~{\rm fb}^{-1}$
collected with the Belle detector~\cite{Belle} at the 
KEKB asymmetric $e^+e^-$ (3.5 on 8~GeV)  collider~\cite{KEKB}.
At KEKB, the $\Upsilon(4S)$ is produced 
with a Lorentz boost of $\beta\gamma=0.425$ along 
the electron beam direction ($z$ direction).
Because the $B^0_d$ and $\overline{B_d}^0$ mesons are nearly at rest in 
the $\Upsilon(4S)$ center of mass system (cms),  
$\Delta t$ can be determined from the $z$ distance
between the $f_{CP}$ and $f_{tag}$ decay vertices,
$\Delta z\equiv z_{CP} - z_{tag}$, as
$\Delta t \simeq \Delta z/\beta\gamma c$.

The Belle detector consists of a three-layer silicon vertex detector
(SVD), a 50-layer central drift chamber (CDC),
an array of 1188 aerogel ${\rm \check{C}}$erenkov counters
(ACC), 128 time-of-flight (TOF) scintillation counters,
and an electromagnetic calorimeter containing 8736 CsI(Tl) 
crystals (ECL) all located 
inside a 3.4-m-diameter superconducting solenoid that generates
a 1.5~T magnetic field.
The transverse momentum resolution for charged tracks is 
$(\sigma_{p_t}/p_t)^2=(0.0019p_t)^2+(0.0034)^2$, where $p_t$ is 
in ${\rm GeV}/c$, and
the impact parameter resolutions
for $p=1~{\rm GeV}/c$ tracks at normal incidence are
$\sigma_{r\phi} \simeq \sigma_{z} = 55 \mu{\rm m}$.
Specific ionization ($dE/dx$) measurements in the CDC
($\sigma_{dE/dx}=6.9\%$ for minimum ionizing pions), 
TOF flight-time measurements
($\sigma_{TOF}=95~{\rm ps}$), and the response of the ACC 
provide $K^\pm$ identification with an
efficiency of $\sim 85\%$ and a charged pion 
fake rate of $\sim 10\%$ for all momenta up to $3.5~{\rm GeV}/c$.
Photons are identified as ECL showers that have a minimum energy of 20 MeV and are not
matched to a charged track. 
The photon energy resolution  is 
$(\sigma_E/E)^2=(0.013)^2+(0.0007/E)^2+(0.008/E^{1/4})^2$,
where $E$ is in GeV.
Electron identification is based on a combination of
CDC $dE/dx$ information,
the ACC response, and the position relative to the extrapolated track, 
shape and energy 
deposit of the associated ECL shower.
The efficiency is greater than $90\%$ 
and the hadron fake rate is  $\sim 0.3\%$ for $p>1~{\rm GeV}/c$. 
An iron flux-return yoke outside the solenoid, 
comprised of 14 layers of 4.7-cm-thick iron 
plates interleaved with a system of resistive plate counters (KLM),
provides muon identification with an efficiency
greater than $90\%$ and a hadron fake rate less than $2\%$
for $p>1~{\rm GeV}/c$.  The KLM is
used in conjunction with the ECL to detect $K_L$ mesons; 
the angular resolution of the $K_L$ direction measurement 
ranges between $1.5^\circ$ and $3^\circ$.

We reconstruct $B^0_d$ decays to the following ${CP}$ eigenstates:
$J/\psi K_S$, $\psi(2S)K_S$, $\chi_{c1}K_S$, $\eta_c K_S$ for $\xi_f=-1$  and
$J/\psi \pi^0$,  $J/\psi K_L$ for $\xi_f=+1$.
The $J/\psi$ and  $\psi(2S)$ mesons are reconstructed via their decays to
$\ell^+\ell^-$ $(\ell=\mu,e)$.
The $\psi(2S)$ is also reconstructed via its  $J/\psi\pi^+\pi^-$ decay,
the $\chi_{c1}$ via its $J/\psi\gamma$ decay, and
the $\eta_c$ via its $K^+K^-\pi^0$
and $K_S(\pi^+\pi^-)K^-\pi^+$~\cite{CC} decays.

For $J/\psi$ and $\psi(2S)\to\ell^+\ell^-$ decays, we
use oppositely charged track pairs where both tracks are 
positively identified as leptons.  For
the $B^0_d\to J/\psi K_S(\pi^+\pi^-)$ mode, the 
requirement for {\em one} of the tracks is relaxed: 
a track with an ECL energy deposit consistent with  
a minimum ionizing particle is accepted
as a muon and a track that satisfies 
either the $dE/dx$ or the ECL shower energy requirements 
as an electron.
For $e^+e^-$ pairs,  we 
include the four-momentum of every photon
detected within 0.05 radians of the
original $e^+$ or $e^-$ direction in the invariant mass calculation.
Nevertheless a 
radiative tail remains and we accept pairs in the asymmetric invariant mass 
interval between $-12.5\sigma$ and $+3\sigma$ of $M_{J/\psi}$ or $M_{\psi(2S)}$,
where $\sigma = 12~{\rm MeV}/c^2$ is the mass resolution.
The $\mu^+\mu^-$ radiative tail is smaller;
we select pairs
within $-5\sigma$ and $+3\sigma$ of $M_{J/\psi}$ or $M_{\psi(2S)}$.
Candidate $K_S\to \pi^+\pi^-$ decays are oppositely 
charged track pairs that have an invariant mass 
within $\pm 4\sigma$ of the $K^0$ mass 
($\sigma\simeq 4~{\rm MeV}/c^2$).
For the $J/\psi K_S$ final state, $K_S\to \pi^0\pi^0$ decays are also used.
For $\pi^0\pi^0$ candidates,
we try all combinations where there are two $\gamma\gamma$ pairs
with an invariant mass  
between 80 and 150~${\rm MeV}/c^2$,
assuming they originate from 
the center of the run-dependent average 
interaction point (IP).
We minimize the sum of the $\chi^2$ values from constrained fits 
of each pair to the $\pi^0$ mass with
$\gamma$ directions determined by
varying the decay point 
along the $K_S$ flight path, which is taken as 
the line from the IP to the energy-weighted center of the four showers.
We select combinations with a $\pi^0\pi^0$ invariant mass
within $\sim \pm 3\sigma$ of $M_{K^0}$, where $\sigma \simeq 9.3~{\rm MeV}/c^2$.
For the $J/\psi\pi^0$ mode, we use a minimum $\gamma$ energy of 100~MeV
and select $\gamma\gamma$ pairs with an invariant mass within
$\pm 3\sigma$ of $M_{\pi^0}$, where 
$\sigma \simeq 4.9~{\rm MeV}/c^2$.

\begin{figure}
\begin{center}
\epsfxsize 5.6 truein \epsfbox{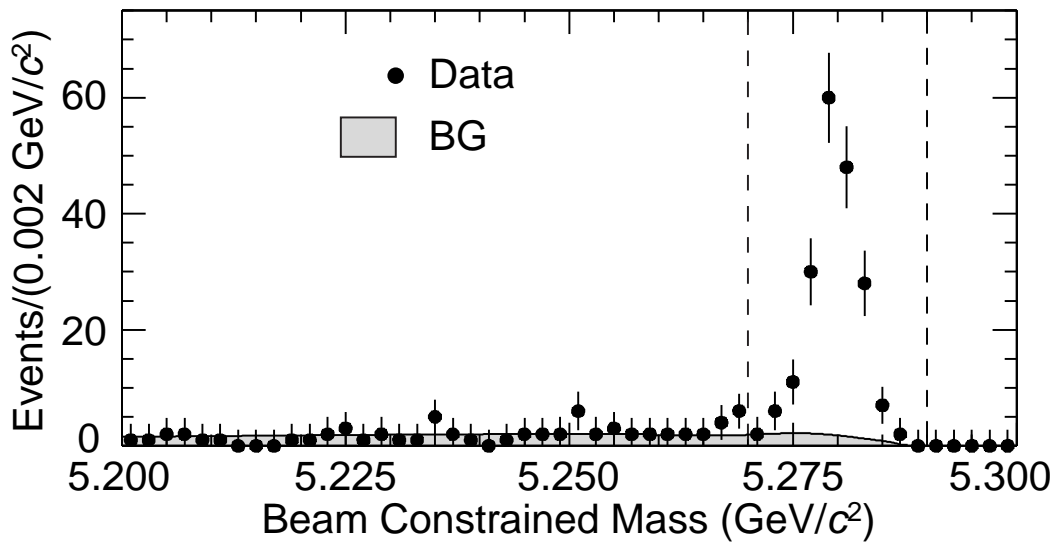}
\end{center}
\caption{The beam-constrained mass distribution for
all decay modes combined (other than  $B^0_d\to J/\psi K_L$).
The shaded area is the estimated background.
The dashed lines indicate the signal region.
}
\label{fig:bmass}
\end{figure}

We isolate reconstructed $B$ meson
decays using 
the energy difference $\Delta E\equiv E_B^{cms} - E_{beam}^{cms}$
and the beam-energy constrained
mass $M_{bc}\equiv\sqrt{(E_{beam}^{cms})^2-(p_B^{cms})^2}$,
where $E_{beam}^{cms}$ is the cms beam energy,
and $E_B^{cms}$ and $p_B^{cms}$ are the cms energy and momentum 
of the $B$ candidate.
Figure~\ref{fig:bmass} shows the $M_{bc}$ 
distribution for 
all channels combined (other than $J/\psi K_L$) after
a  $\Delta E$ selection that varies 
from $\pm 25$~MeV to $\pm 100$~MeV (corresponding to $\sim \pm 3\sigma$), 
depending on the mode.
The $B$ meson signal region is defined as 
$5.270<M_{bc}<5.290~{\rm GeV}/c^2$;
the $M_{bc}$ resolution is $3.0~{\rm MeV}/c^2$.
Table~\ref{tab:tally} lists
the numbers of  observed events ($N_{ev}$) and
the background ($N_{bkgd}$) determined by
extrapolating the event rate in the
non-signal  $\Delta E$ {\em vs.}~$M_{bc}$ region 
into the signal region.

\begin{figure}
\begin{center}
\epsfxsize 5.6 truein \epsfbox{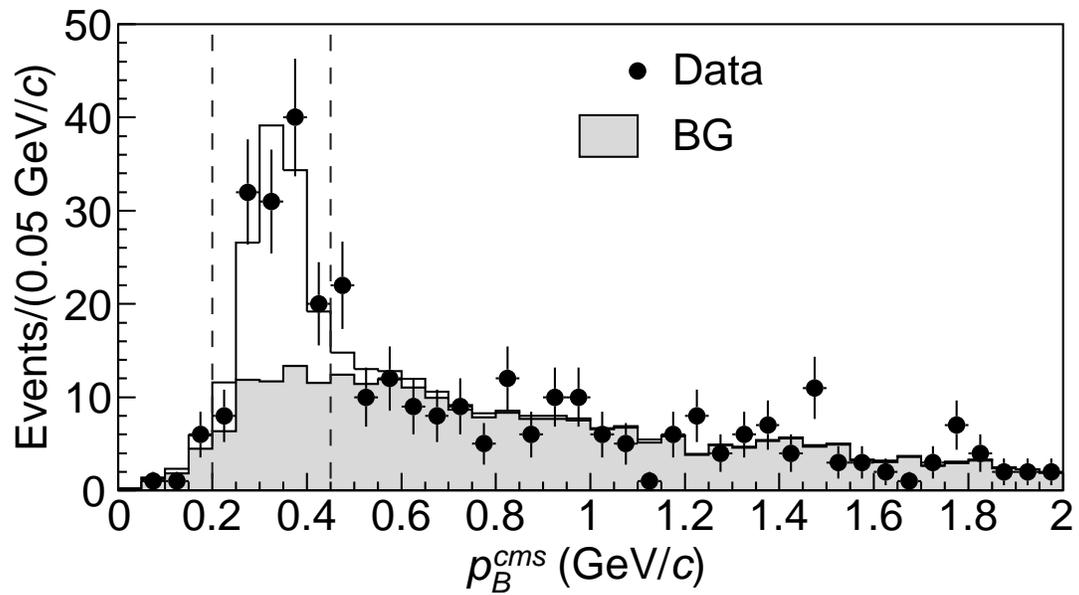}
\end{center}
\caption{The $p_B^{cms}$ distribution for $B^0_d\to J/\psi K_L$ candidates
with the results of the fit.
The solid line is the signal plus background;
the shaded area is background only.
The dashed lines indicate the signal region.
}
\label{fig:pbstar}
\end{figure}

Candidate $B^0_d\to J/\psi K_L$  decays are selected by requiring 
the observed $K_L$ direction
to be within $45^\circ$ from the direction 
expected for a two-body decay
(ignoring the $B_d^0$ cms motion).
We reduce the background by means of
a likelihood quantity that 
depends on the $J/\psi$ cms momentum, 
the  angle between the $K_L$ and its nearest-neighbor charged track, 
the charged track multiplicity, and 
the kinematics that obtain when the event is reconstructed
assuming a $B^+ \to$ $J/\psi K^{*+}(K_L\pi^+)$ hypothesis.
In addition, we remove events
that are reconstructed as 
$B_d^0 \to J/\psi K_S$, $J/\psi K^{*0}(K^+\pi^-, K_S \pi^0)$,
$B^+\to$ $J/\psi K^+$, or  $J/\psi K^{*+}(K^+ \pi^0$ $K_S \pi^+)$ decays.
Figure~\ref{fig:pbstar} shows the $p_B^{cms}$ distribution,
calculated for a $B^0_d \to J/\psi K_L$ two-body decay hypothesis,
for the surviving events.
The histograms in the figure are the  results of a fit to the signal
and background distributions,  where the shapes are
derived from Monte Carlo simulations (MC)~\cite{MC},
and  the 
normalizations are allowed to vary.
Among the total of 131 entries in the 
$0.2\leq p_B^{cms}\leq 0.45~{\rm GeV}/c$ signal region, the fit finds
77 $J/\psi K_L$ events.

The leptons and charged pions and kaons
among the tracks that are not associated with $f_{CP}$  
are used to identify
the flavor of the accompanying $B$ meson.
Tracks are selected in several categories
that distinguish 
the $b$-flavor by the track's charge:  
high momentum leptons 
from  $b\to c\ell^-\overline{\nu}$,
lower momentum leptons from  $c\to s\ell^+\nu$,
charged kaons from $b\to c\to s$,
high momentum pions from decays of the type 
$B_d^0\to D^{(*)-}(\pi^+, \rho^+, a_1^+, {\rm etc.})$, and
slow pions from $D^{*-}\to \overline{D}^0\pi^-$.
For each track 
in one of these categories, 
we use the MC to
determine the relative probability that it
originates from a $B^0_d$ or $\overline{B_d}^0$ 
as a function of
its charge, 
cms momentum and polar angle,
particle-identification probability, and other kinematic and
event shape quantities.
We combine the results from the different track categories
(taking into account correlations for the case of multiple inputs)
to determine a $b$-flavor $q$, where
$q = +1$  when $f_{tag}$  is more likely to be a $B^0_d$
and $-1$ for a $\overline{B_d}^0$.  We use the MC to evaluate
an event-by-event flavor-tagging dilution factor, $r$, which
ranges from $r=0$ for no flavor discrimination to $r=1$ for
perfect flavor assignment.
We only use $r$ to categorize the event.
For the $CP$ asymmetry analysis, we use the data to correct for 
wrong-flavor assignments.

The probabilities for an 
incorrect flavor assignment, $w_l\ (l=1,6)$,
are measured directly from the data for six  $r$ intervals
using a sample of exclusively reconstructed, self-tagged
$B^0_d\to D^{*-}\ell^+\nu$, $D^{(*)-}\pi^+$, 
and $D^{*-}\rho^+$ decays.
The $b$-flavor of the accompanying $B$ meson
is assigned according to the above-described flavor-tagging algorithm,
and values of
$w_l$ are determined from the amplitudes of the 
time-dependent $B^0_d$-$\overline{B_d}^0$ mixing oscillations~\cite{mixing}:
$(N_{OF} - N_{SF})/(N_{OF}+N_{SF})
=(1-2w_l )\cos (\Delta m_d \Delta t)$.
Here $N_{OF}$ and $N_{SF}$ are the numbers of opposite and same
flavor events.
Table~\ref{tab:tag} lists the resulting  $w_l$ values 
together with the fraction of the events ($f_l$)
in each $r$ interval.
All events in Table~\ref{tab:tally} fall in one of the six $r$ intervals.
The total effective tagging efficiency is 
$\sum_l f_l(1-2w_l)^2 = 0.270^{+0.021}_{-0.022}$, 
where the error includes both statistical and systematic uncertainties,
in good agreement  with the MC result of 0.274.
We check for a possible bias in the flavor tagging 
by measuring the effective tagging efficiency 
for $B_d^0$ and $\overline{B_d}^0$ self-tagged samples separately,
and  for different  $\Delta t$ intervals.
We find no statistically significant difference.

The vertex positions for the $f_{CP}$ and $f_{tag}$ decays are 
reconstructed using tracks
that have at least one 
3-dimensional coordinate determined from associated $r\phi$ and $z$
hits in the same SVD layer 
plus one or more additional $z$ hits in other SVD layers.  
Each vertex position is required to be 
consistent with the IP profile smeared in the
$r\phi$ plane by the $B$ meson decay length.
(The IP size,
determined run-by-run, is typically $\sigma_{x}\simeq 100~\mu{\rm m}$,
$\sigma_{y}\simeq 5~\mu{\rm m}$ and $\sigma_{z}\simeq 3~{\rm mm}$.)
The $f_{CP}$ vertex is determined using 
lepton tracks  from 
the $J/\psi$ or $\psi(2S)$ decays, or prompt tracks from $\eta_c$ decays. 
The $f_{tag}$ vertex 
is determined from tracks not assigned to $f_{CP}$
with additional requirements 
of  $\delta r<0.5~{\rm mm}$, $\delta z<1.8~{\rm mm}$ and
$\sigma_{\delta z}<0.5~{\rm mm}$,
where $\delta r$ and $\delta z$ are the distances 
of the closest approach to the $f_{CP}$ vertex 
in the $r\phi$ plane and the $z$ direction,
respectively, and $\sigma_{\delta z}$ is the calculated error of $\delta z$.
Tracks that form a $K_S$ are removed. 
The MC indicates that the average $z_{CP}$ resolution 
is  $75~\mu{\rm m}~({\rm rms})$;
the $z_{tag}$ resolution is worse ($140~\mu{\rm m}$) 
because of the lower average momentum of the $f_{tag}$ decay products 
and the smearing caused by secondary tracks from charmed meson decays.

The resolution function $R(\Delta t)$ for the proper time interval is parameterized 
as a sum of two  Gaussian components: a {\it main} component 
due to  the SVD vertex resolution, charmed meson lifetimes
and the effect of the cms motion of the $B$ mesons,
plus a {\it tail} component caused by poorly reconstructed tracks.
The means 
($\mu_{main}$, $\mu_{tail}$) 
and widths 
($\sigma_{main}$, $\sigma_{tail}$)
of the Gaussians are
calculated event-by-event from the 
$f_{CP}$ and $f_{tag}$ vertex fit error matrices;
average values are $\mu_{main}=-0.09~{\rm ps}$, $\mu_{tail}=-0.78~{\rm ps}$ and
$\sigma_{main}=1.54~{\rm ps}$, $\sigma_{tail}=3.78~{\rm ps}$. 
The negative values of the means are
due to secondary tracks from charmed mesons.
The relative fraction of the main Gaussian 
is determined to be $0.982\pm 0.013$ from a
study of $B^0_d\to D^{*-}\ell^+\nu$ events.
The reliability of the $\Delta t$ 
determination and $R(\Delta t)$ parameterization is confirmed
by lifetime measurements of
the neutral and charged $B$ mesons~\cite{tajima} 
that use the same procedures and are in good agreement with  
the world average values~\cite{PDG}.

We determine $\sin 2\phi_1$ from an 
unbinned maximum-likelihood fit to the observed $\Delta t$ distributions.
The probability density function (pdf) expected for the 
signal distribution is given by
\begin{eqnarray*}
{\cal P}_{sig}(\Delta t,q,w_l,\xi_f)
&=&\frac{e^{-|\Delta t|/\tau_{B^0_d}}}{2\tau_{B^0_d}}%
\{1-\xi_f q(1-2w_l)\sin 2\phi_1\sin (\Delta m_d\Delta t )\},
\end{eqnarray*}
where 
we fix the $B^0_d$ lifetime and  mass difference 
at their world average values~\cite{PDG}.
The pdf used for background events is
${\cal P}_{bkg}(\Delta t)=f_\tau e^{-|\Delta t|/\tau_{bkg}}/2\tau_{bkg}+(1-f_\tau)\delta(\Delta t),$
where $f_\tau$ is the fraction of the background component 
with an effective lifetime $\tau_{bkg}$ and $\delta(\Delta t)$ is the Dirac delta function.
For all $f_{CP}$ modes except $J/\psi K_L$ we find 
$f_\tau=0.10^{+0.11}_{-0.05}$ and
$\tau_{bkg}=1.75^{+1.15}_{-0.82}~{\rm ps}$ 
using events in background-dominated regions of $\Delta E$ {\em vs.}~$M_{bc}$.
The $J/\psi K_L$  background is dominated by $B \to J/\psi X$ decays,
where some final states are
$CP$ eigenstates and need special treatment.
A MC study shows that
the background contribution from  the $\xi_f=-1$ sources $J/\psi K_S$,
$\psi(2S)K_S$ and $\chi_{c1}K_S$   
is 7.9\%, while that from the $\xi_f=+1$
$\psi(2S)K_L$ and $\chi_{c1}K_L$  modes is $7.0\%$. 
Thus, the effects on the $CP$ asymmetry from these states nearly cancel.
The remaining dominant $CP$ mode, $J/\psi K^*(K_L\pi^0)$, 
which accounts for 19\% of the
total background, is taken to be a $73/27$ mixture
of $\xi_f=-1$ and $+1$, respectively, based on our measurement of the $J/\psi$
polarization in the $B_d^0\to J/\psi K^{*0}(K_S\pi^0)$ decay~\cite{paper285}.
For the $J/\psi K^*(K_L\pi^0)$ background pdf  we use ${\cal P}_{sig}$ with effective $CP$ eigenvalue
$\xi_f=-0.46^{+1.46}_{-0.54}$,
where the error has been expanded to include all possible values.
For the non-$CP$ background modes 
we use ${\cal P}_{bkg}$ with $f_{\tau}=1$ and $\tau_{bkg}=\tau_B$.
%

\begin{figure}
\begin{center}
\epsfxsize 5.6 truein \epsfbox{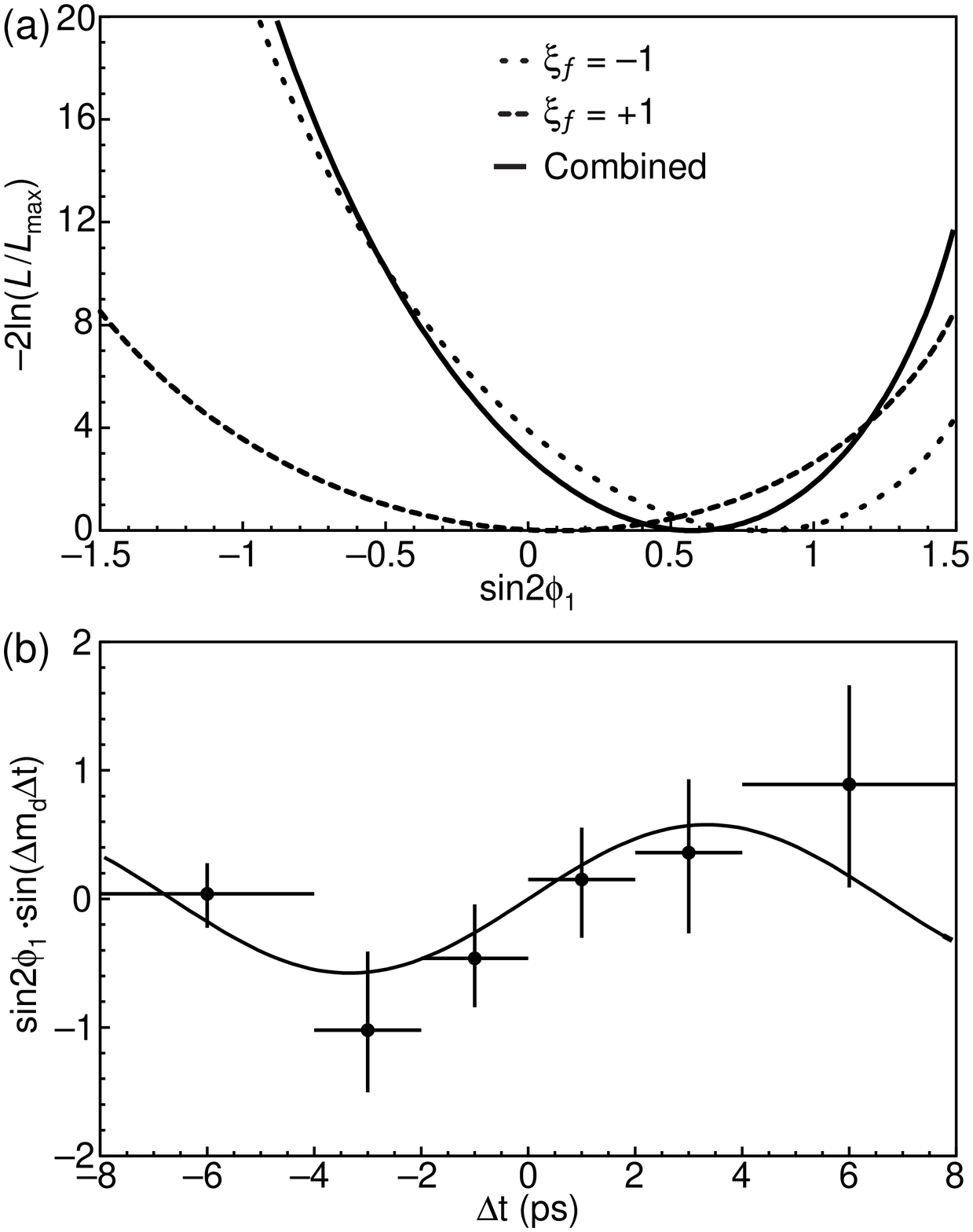}
\end{center}
\caption{
(a) Values of $-2\ln(L/L_{max})$ 
{\em vs.}~$\sin 2\phi_1$ for the $\xi_f=-1$ and  $+1$ modes separately and 
for both modes combined. 
(b) The asymmetry obtained 
from separate fits to each $\Delta t$ bin;
the curve is the result of the global fit ($\sin 2\phi_1=0.58$).
}
\label{fig:cpfit}
\end{figure}

The pdfs are convolved with $R(\Delta t)$ to determine
the likelihood value for each event as a function of $\sin 2\phi_1$:
\begin{eqnarray*}
{\cal L}_i&=&{\displaystyle \int}\{f_{sig}{\cal P}_{sig}(\Delta t^\prime, q,w_l,\xi_f)
+(1-f_{sig}){\cal P}_{bkg}(\Delta t^\prime)\} R(\Delta t-\Delta t^\prime)d\Delta t^\prime,
\end{eqnarray*}
where $f_{sig}$ is the probability that the event is signal,
calculated as a function of $p_B^{cms}$ for $J/\psi K_L$ and
of $\Delta E$ and $M_{bc}$ for other modes.
The most probable $\sin 2\phi_1$ is the
value that maximizes 
the likelihood function 
$L=\prod_i {\cal L}_i$, where the product is over all
events.
We performed a blind analysis:
the fitting algorithms were developed and finalized using a flavor-tagging routine
that does not divulge the sign of $q$.
The sign of $q$ was then turned on and the application of 
the fit to all the events listed in Table~\ref{tab:tally}
produces the result  $\sin2\phi_1=0.58^{+0.32+0.09}_{-0.34-0.10}$,
where the first error is statistical
and the second systematic. 
The systematic errors are  dominated by 
the uncertainties in  $w_l$ ($^{+0.05}_{-0.07}$) and
the $J/\psi K_L$ background ($\pm 0.05$).
Separate fits to the $\xi_f=-1$ and  $\xi_f = +1$ event samples
give  $0.82^{+0.36}_{-0.41}$ and $0.10^{+0.57}_{-0.60}$, respectively~\cite{fits}.  
Figure~\ref{fig:cpfit}(a) shows $-2\ln(L/L_{max})$ 
as  a function of 
$\sin 2\phi_1$ for the $\xi_f=-1$ and  $\xi_f=+1$ modes separately 
and for both modes combined. 
Figure~\ref{fig:cpfit}(b) shows the 
asymmetry 
obtained by performing the fit to events in  $\Delta t$ bins separately,
together with a curve that 
represents $\sin 2\phi_1\sin (\Delta m_d\Delta t)$ 
for $\sin 2\phi_1=0.58$.

We check for a possible
fit bias by applying the same fit to  non-$CP$
eigenstate modes: $B^0_d \rightarrow D^{(*)-}\pi^+$, $D^{*-}\rho^+$,
$J/\psi K^{*0}(K^+\pi^-)$, and $D^{*-}\ell^+\nu$,
where ``$\sin 2\phi_1$'' should be zero, and the charged mode $B^+ \to J/\psi K^+$.
For all the modes combined we find $0.065\pm 0.075$, consistent with a null asymmetry.

We have presented a measurement of the 
Standard Model $CP$ violation parameter
$\sin 2\phi_1$ based on a $10.5~{\rm fb}^{-1}$ data sample collected at
the $\Upsilon(4S)$:
$$\sin 2\phi_1=0.58^{+0.32}_{-0.34}({\rm stat})^{+0.09}_{-0.10}({\rm syst}).$$
The probability of observing $\sin 2\phi_1>0.58$ if the true value is zero
is  $4.9\%$.
Our measurement is more precise than 
the previous measurements~\cite{old1}
and consistent with
SM constraints~\cite{SMpred}.

We wish to thank the KEKB accelerator group 
for the excellent operation.
We acknowledge support from the Ministry of Education, Culture, Sports, Science and
Technology of Japan and
the Japan Society for the Promotion of Science;
the Australian Research Council and the Australian Department of Industry,
Science and Resources;
the Department of Science and Technology of India;
the BK21 program of the Ministry of Education of Korea and
the SRC program of the Korea Science and Engineering Foundation;
the Polish State Committee for Scientific Research 
under contract No.2P03B 17017; 
the Ministry of Science and Technology of Russian Federation;
the National Science Council and the Ministry of Education of Taiwan;
the Japan-Taiwan Cooperative Program of the Interchange Association;
and the U.S. Department of Energy.

\newpage
\begin{table}
\caption{The numbers of $CP$ eigenstate events}
\label{tab:tally}
\begin{tabular}{lrr}
Mode & $N_{ev}$ & $N_{bkgd}$\\
\hline
$J/\psi(\ell^+\ell^-) K_S(\pi^+\pi^-)$ & 123 & 3.7\\
$J/\psi(\ell^+\ell^-) K_S(\pi^0\pi^0)$  & 19 & 2.5\\
$\psi(2S)(\ell^+\ell^-)K_S(\pi^+\pi^-)$  & 13 & 0.3\\
$\psi(2S)(J/\psi\pi^+\pi^-)K_S(\pi^+\pi^-)$ & 11 & 0.3\\
$\chi_{c1}(\gamma J/\psi) K_S(\pi^+\pi^-)$ & 3 & 0.5\\
$\eta_c(K^+K^-\pi^0)K_S(\pi^+\pi^-)$ & 10 & 2.4\\
$\eta_c(K_S K^+\pi^-)K_S(\pi^+\pi^-)$ & 5& 0.4\\
$J/\psi(\ell^+\ell^-) \pi^0$ & 10 & 0.9 \\
\hline
Sub-total & 194 & 11    \\ 
\hline
$J/\psi(\ell^+\ell^-) K_L$ & 131 & 54 
\end{tabular} 
\end{table}

\begin{table}
\caption{
Experimentally determined
event fractions ($f_l$) 
and incorrect flavor assignment probabilities ($w_l$)
for each $r$ interval.
}
\label{tab:tag}
\begin{tabular}{lccc}
$l$&$r$ & $f_l$ & $w_l$\\
\hline
1&$0.000-0.250$ & $0.393\pm 0.014$ & $0.470^{+0.031}_{-0.035}$ \\
2&$0.250-0.500$ & $0.154\pm 0.007$ & $0.336^{+0.039}_{-0.042}$ \\
3&$0.500-0.625$ & $0.092\pm 0.005$ & $0.286^{+0.037}_{-0.035}$\\
4&$0.625-0.750$ & $0.100\pm 0.005$ & $0.210^{+0.033}_{-0.031}$\\
5&$0.750-0.875$ & $0.121\pm 0.006$ & $0.098^{+0.028}_{-0.026}$\\
6&$0.875-1.000$ & $0.134\pm 0.006$ & $0.020^{+0.023}_{-0.019}$
\end{tabular}
\end{table}

%
%

%
%


\begin{references}

\bibitem{KM}
M.~Kobayashi and T.~Maskawa, Prog. Theor. Phys. {\bf 49}, 652 (1973).

\bibitem{carter}
A.B.~Carter and A.I.~Sanda, Phys. Rev. {\bf D23}, 1567 (1981);
I.I.~Bigi and A.I.~Sanda, Nucl. Phys. {\bf B193}, 85 (1981).

\bibitem{Sanda} 
H.~Quinn and A.I.~Sanda, Eur. Phys. Jour. {\bf C15}, 626 (2000).
(Some papers refer to this angle as $\beta$.)

\bibitem{Belle}
K.~Abe {\it et al.} (Belle Collab.), 
{\em The Belle Detector}, KEK Report 2000-4, to be published
in Nucl. Instrum. Methods.

\bibitem{KEKB}
KEKB B Factory Design Report, KEK Report 95-1, 1995, unpublished.


\bibitem{CC}
Throughout this Letter, when a mode is quoted the inclusion of the charge 
conjugate mode is implied.


\bibitem{MC}
We use the QQ $B$ meson decay event generator 
developed by the CLEO collaboration (http://www.lns.cornell.edu%
/public/CLEO/soft/QQ)
and GEANT3 for the detector simulation;
CERN Program Library Long Writeup W5013, CERN, 1993.

\bibitem{mixing}
J.~Suzuki (Belle Collab.), {\em Determination of $B^0$-$\overline{B}^0$ Mixing from The Time
Evolution of Dilepton and $B^0\to D^{(*)-}\ell^+\nu$ Events at Belle},
Proceedings of the 30th International Conference on High Energy 
Physics, July 2000, Osaka. 

\bibitem{tajima}
H.~Tajima (Belle Collab.), {\em Measurement of Heavy Meson 
Lifetimes with Belle}, {\it ibid.}


\bibitem{PDG}
D.E.~Groom {\it et al.} (PDG), Eur. Phys. Jour.  {\bf C15}, 1 (2000).

\bibitem{paper285}
Belle Collab., {\em Measurement of Polarization of $J/\psi$ in 
$B^0\to$ $J/\psi +K^{*0}$ and
$B^+\to$ $J/\psi+ K^{*+}$ Decays}, Contributed paper (\#285) to
the 30th International Conference on High Energy Physics, 
July 2000, Osaka.
This result agrees within errors with those of
C.P~Jessop {\it et al.} (CLEO Collab.), Phys. Rev. Lett. {\bf 79}, 4533 (1997) and 
T.~Affolder {\it et al.} (CDF Collab.), Phys. Rev. Lett. {\bf 85}, 4668 (2000).

\bibitem{fits}
A fit to only the $B^0_d\to J/\psi K_S(\pi^+\pi^-)$ events
gives a $\sin 2\phi_1$ value of $1.21^{+0.40}_{-0.47}$;
a fit to only the non-$J/\psi K_S$ $\xi=-1$ modes gives $-0.05^{+0.76}_{-0.74}$. 
Separate fits to the $q=+1$ and $q=-1$ event samples
give $\sin2\phi_1$ values of $0.40^{+0.47}_{-0.49}$ and $0.73^{+0.41}_{-0.46}$,
respectively.

\bibitem{old1}
K.~Ackerstaff {\it et al.} (OPAL Collab.), Eur. Phys. Jour. {\bf C5}, 379 (1998);
T.~Affolder {\it et al.} (CDF Collab.), Phys. Rev. {\bf D61}
072005 (2000); and
R.~Barate {\it et al.} (ALEPH Collab.), Phys. Lett. {\bf B492}, 259 (2000).

\bibitem{SMpred}
For example: S.~Mele, Phys. Rev. {\bf D59}, 113011 (1999).

\end{references}
\end{document}